\documentclass[lettersize,journal]{IEEEtran}
\usepackage{amsmath,amssymb,amsfonts}
\usepackage{algorithmic}
\usepackage{algorithm}
\usepackage{array}
\usepackage[caption=false,font=footnotesize,labelfont=rm,textfont=rm]{subfig}
\usepackage{textcomp}
\usepackage{stfloats}
\usepackage{url}
\usepackage{verbatim}
\usepackage{graphicx}
\usepackage{cite}
\usepackage{xcolor}
\usepackage{subfig}
\usepackage{titlesec}
\usepackage{mathrsfs}
\usepackage{bm,bbold}
\newcommand\bw{\ensuremath{{\bm w}}}
\newcommand\tw{\ensuremath{{\widetilde \bw}}}

\newcommand\bq{\ensuremath{{\bm q}}}
\newcommand\bx{\ensuremath{{\bm x}}}
\newcommand\by{\ensuremath{{\bm y}}}

\newcommand\bh{\ensuremath{{\bm h}}}

\newcommand\bz{\ensuremath{{\bm z}}}

\newcommand\bR{\ensuremath{{\bold R}}}

\newcommand\bt{\ensuremath{{\bm t}}}

\newcommand\bC{\ensuremath{{\bold C}}}

\newcommand\ba{\ensuremath{{\bm a}}}
\newcommand\bda{\ensuremath{{\bold a}}}

\newcommand\bA{\ensuremath{{\bold A}}}

\newcommand\bb{\ensuremath{{\bm b}}}

\newcommand\bg{\ensuremath{{\bm g}}}

\newcommand\bd{\ensuremath{{\bm d}}}
\newcommand\bD{\ensuremath{{\bold D}}}
\newcommand\bu{\ensuremath{{\bm u}}}

\newcommand{\Rbb}{\mathbb{R}}
\newcommand{\Cbb}{\mathbb{C}}

\newcommand{\setX}{\mathcal{X}}

\newcommand{\setC}{\mathcal{C}}
\newcommand{\setN}{\mathcal{N}}

\newcommand{\Exp}{\mathbb{E}}

\newcommand{\diag}{\mathrm{diag}}

\newcommand\br{\ensuremath{{\bm r}}}

\newcommand{\bI}{{\bold I}}

\begin{document}

\title{An Efficient Sum-Rate Maximization Algorithm for Fluid Antenna-Assisted ISAC System}

\author{Qian Zhang, \IEEEmembership{Graduate Student Member,~IEEE}, Mingjie Shao, \IEEEmembership{Member, IEEE}, Tong Zhang, \IEEEmembership{Member, IEEE},\\ Gaojie Chen, \IEEEmembership{Senior Member, IEEE}, Ju Liu, \IEEEmembership{Senior Member, IEEE}, and P. C. Ching, \IEEEmembership{Life Fellow, IEEE}
\thanks{
	This work was supported in part by the National Natural Science Foundation of China under Grant 62071275, 62401340, and 62401229; in part by the Natural Science Foundation of Shandong Province under Grant ZR2023QF103; in part by the Guangdong Basic and Applied Basic Research Project under Grant 2023A1515110477.
	Corresponding authors: Ju Liu and Mingjie Shao.}
\thanks{Qian Zhang, Mingjie Shao, and Ju Liu are with School of Information Science and Engineering, Shandong University, Qingdao 266237, China (qianzhang2021@mail.sdu.edu.cn; \{mingjieshao, juliu\}@sdu.edu.cn).}
\thanks{Tong Zhang is with the Institute of Intelligent Ocean Engineering, Harbin Institute of Technology, Shenzhen, 518055, China (tongzhang@hit.edu.cn).}
\thanks{Gaojie Chen is with the School of Flexible Electronics (SoFE) \& State Key Laboratory of Optoelectronic Materials and Technologies (OEMT), Sun Yat-sen University, Shenzhen, 518107, China (e-mail: gaojie.chen@ieee.org).}
\thanks{P. C. Ching is with the Department of Electronic Engineering, The Chinese University of Hong Kong, Hong Kong (e-mail: pcching@ee.cuhk.edu.hk). }
}

\markboth{}
{Shell \MakeLowercase{\textit{et al.}}: Bare Demo of IEEEtran.cls for IEEE Journals}
\maketitle

\begin{abstract}
This letter investigates a fluid antenna (FA)-assisted integrated sensing and communication (ISAC) system, with joint antenna position optimization and waveform design. We consider enhancing the sum-rate maximization (SRM) and sensing performance with the aid of FAs. Although the introduction of FAs brings more degrees of freedom for performance optimization, its position optimization poses a non-convex programming problem and brings great computational challenges. This letter contributes to building an efficient design algorithm by the block successive upper bound minimization and majorization-minimization principles, with each step admitting closed-form update for the ISAC waveform design. In addition, the extrapolation technique is exploited further to speed up the empirical convergence of FA position design. Simulation results show that the proposed design can achieve state-of-the-art sum-rate performance with at least 60\% computation cutoff compared to existing works with successive convex approximation (SCA) and particle swarm optimization (PSO) algorithms.

\end{abstract}

\vspace{-0.1cm}

\IEEEpeerreviewmaketitle

\vspace{-0.3cm}
\section{Introduction}
\IEEEPARstart{W}{ith} the development of wireless networks, a large number of emerging applications, such as vehicle-to-everything (V2X), demand for joint communication and sensing capabilities in wireless communication systems~\cite{Cui2021ISAC,Liu2018ISAC,Dong2023R_ISAC}.
Consequently, integrated sensing and communication (ISAC) has been proposed to allow communications and radar sensing to share the same frequency and hardware resources, thus improving spectral, energy, and hardware efficiency~\cite{Liu2018DFRC,Liu2022ISAC,Zhang2023ISAC_EI}.
However, the concept of ISAC increases the resource demands required to effectively fulfill both.

Recently, fluid antennas (FAs) have been proposed as a novel technique to further improve the spectral efficiency of the MIMO systems \cite{Wong2023FA}.
The increasing body of studies has shown that the optimized FA positions can benefit the communication performance.
In \cite{Xiao2024CEMA} and \cite{New2024FAChannelEstimation_2}, the authors proposed a method that can reconstruct the full CSI corresponding to all FA position combinations from the estimated CSI under the partial FA position combinations. 
In addition, the joint design with FA positions requires dedicated optimization techniques to improve the spectral efficiency \cite{Wong2023FA} or ISAC performance~\cite{Wang2024FA_ISAC,Zou2024FA_ISAC}. 
In~\cite{Wang2024FA_ISAC}, deep reinforcement learning (DRL) was adopted to address FA-assisted ISAC problems, while DRL consumes a large amount of training data and time.
In~\cite{Zou2024FA_ISAC}, the FAs with predetermined discrete positions were considered to balance the ISAC performance, and effective FA ports were activated by sparsity constraints.
Moreover, movable antennas (MAs) with the same function as the FAs were proposed to enhance the ISAC system~\cite{Kuang2024MA_ISAC,Wu2024MA_ISAC}.
In~\cite{Kuang2024MA_ISAC}, the particle swarm optimization (PSO) algorithm was employed to design the MA positions for achieving better ISAC, while the PSO algorithm may have high computational overhead and the performance can be sensitive to the number of particles and searches.
In~\cite{Wu2024MA_ISAC}, a reconfigurable intelligent surface (RIS) was used to assist the MA system in sensing non-line-of-sight targets, and the optimized MA positions were achieved by the successive convex approximation (SCA) algorithm.
While the existing works have demonstrated the potential improvement of considering FAs in ISAC, the challenge of the high computational complexity of joint design remains a key issue, as also pointed out in \cite{Wong2023FA_PartII}.


This letter aims to alleviate the design computational complexity in FA-aided ISAC systems under a proposed formulation for optimized sum-rate performance under sensing requirements. The development is based on the majorization minimization (MM) principle, where a novel proximal distance algorithm (PDA) is proposed to obtain the closed-form beamformer at each iteration, and an extrapolated projected gradient (EPG) method is proposed to speed up the optimization of antenna positions. Compared to the existing widely adopted PSO, our approach makes more use of the problem structure and shows improved performance and reduced complexity. Simulations demonstrate that our algorithm achieves top-tier sum-rate performance with at least 60\% less computation compared to existing methods like SCA and PSO. For larger FA scales, it provides over 40\% sum-rate improvement compared to the SCA method, with significant computational savings and over 150 times faster performance than the PSO method. Additionally, incorporating FA results in a more than 30\% performance boost compared to traditional ISAC without FA.

\vspace{-0.3cm}
\section{System Model}
\begin{figure}[t]	
	\centering \includegraphics[width=0.75\linewidth]{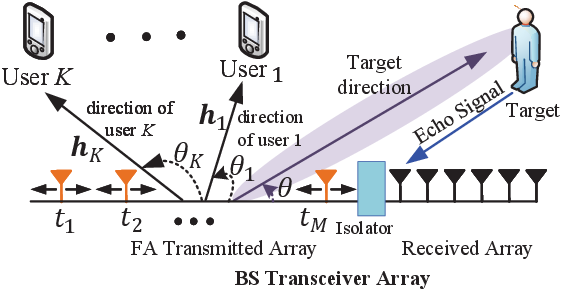}
	\vspace{-0.3cm}
	\caption{The model of fluid antenna-assisted ISAC system.}
	\label{fig:system_model}
	\vspace{-0.5cm}
\end{figure}
We consider a downlink transmission in an FA-assisted ISAC system, as shown in Fig.~\ref{fig:system_model}.
The BS equipped with $M$ FAs communicates with $K$ single-antenna users and senses a point-like target at the same time.
The target echo signal can be received by a specialized received antenna, cf. \cite{Liu2018ISAC,Dong2023R_ISAC}, in which an isolator exists between the transmitted antenna and the received antenna, so there is no interference.
We assume that FAs can move in a linear array of length $D$.
We denote $t_m \in [0,D]$ as the position of the $m$th FA and the assembled vector $\bt = [t_1,t_2,\dots,t_M] \in \mathbb{R}^{M}$ as the antenna position vector (APV)\footnote{For the sake of exposition, we consider one-dimensional (1D) FA arrays. However, the proposed beamforming method can be directly extended to two-dimensional (2D) or three-dimensional (3D) FA arrays by alternately optimizing single FA position.}, where APV can be configured with a high refresh rate \cite{Wong2023FA}.
In such an ISAC system, the BS transmits a dual-functional signal for sensing and communication under block fading channels, where the BS-user channels are modeled by the following line-of-sight (LoS) model.
\begin{equation}
	\begin{split}
		\bh_k = \delta_k \bda(\bt,\theta_k), ~ k \in {\cal K} \triangleq \{1,2,\dots,K\},
	\end{split}
\end{equation}
where 
$
\bda(\bt,\theta_k) = \left[ e^{j v_k t_1}, e^{j v_k t_2}, \dots, e^{j v_k t_M }  \right]^{\rm T}
$; $v_k = \frac{2\pi}{\lambda}{\rm cos}(\theta_k)$; $\lambda$ is the wavelength; $\theta_k$ is the angle of departure (AOD) of the FA array along the direction of the user $k$; and $\delta_k$ denotes the propagation gain.

We employ the linear beamformer to transmit the dual-functional signal, which is given by $\bx_t = \sum_{k=1}^{K} \bw_k s_k ,$
where $\bw_k \in \mathbb{C}^{M\times 1}$ is a linear beamformer for user $k$; $s_k$ is the data symbol for user $k$; and $s_k \sim {\cal CN}(0,1) $.
Then, the $k$th user's received signal is given by
\begin{equation}
	\begin{split}
		y_k = \bh_k^{\rm H} \bx_t + n_k,~k \in {\cal K},
	\end{split}
\end{equation}
where $n_k\sim \setC\setN(0,\sigma_k^2)$ denotes the noise at user $k$.
The signal-to-interference-plus-noise ratio (SINR) of user~$k$ is denoted as
\begin{equation}
	\begin{split}
		\gamma_k = \frac{ |\bh_k^{\rm H} \bw_k|^2  }{\sum_{i=1,i\neq k}^{K} |\bh_k^{\rm H} \bw_i|^2 + \sigma_k^2 }, ~ k \in {\cal K}.
	\end{split}
\end{equation}

In the considered ISAC system, the transmitted signal can also detect the presence or state of a target in the interested direction~\cite{Liu2018DFRC,Stoica2007Radar}.
This requires the ISAC system to provide sufficient pulse power along the probed direction.
To describe this, the covariance matrix of $\bx_t$ is given by
\[
	\bR_w = \Exp[\bx_t\bx_t^{\rm H}] =  \sum_{k=1}^{K} \bw_k \bw_k^{\rm H}.
\]
We let the probing power along the interested direction greater than a threshold, which is posed as~\cite{Liu2018DFRC,Liu2022ISAC}
\begin{equation} \label{probing_power}
	\begin{split}
		P(\bw,\bt,\theta) = \bda^{\rm H}(\bt,\theta) \, \bR_w \, \bda(\bt,\theta)\geq P_t,
	\end{split}
\end{equation}
where $\bw = [\bw_1^{\rm T},\dots,\bw_M^{\rm T}]^{\rm T}$; $\theta$ is the AOD of the FA array along the probed direction.

\section{Beamformer and APV Optimization}

We aim to optimize the beamformer and APV to maximize the multiuser sum rate while the probing power in the target direction is above a prefixed level.
The problem formulation of the joint design is
\begin{subequations}\label{op:pri_prob}
	\begin{align}
		&\max_{ \bw,\bt }\,\,\,
		\sum_{k=1}^{K} {\rm log}_2\, (1 + \gamma_k)  \\[-3pt]
		\mbox{s.t. }
		&~ \setC_{\sf BS}: \sum_{k=1}^{K} \Vert \bw_k \Vert_2^2 \leq P_{max} , \label{eq:const1}\\[-3pt]
		&~ \setC_{t}\,\,\,: P(\bw,\bt,\theta) \geq P_t, \label{eq:const2} \\[-3pt]
		&~ \setC_{\sf FR}: t_1 \geq 0, ~ t_M \leq D, \label{eq:const3} \\[-3pt]
		&~ \setC_{\sf AC}: t_m-t_{m-1}\geq D_0,~ m =2,3,\dots,M. \label{eq:const4}
	\end{align}
\end{subequations}
In problem \eqref{op:pri_prob}, the constraint~\eqref{eq:const1} portrays the power budget at the BS;
constraints~\eqref{eq:const3} and~\eqref{eq:const4} restrict the FA positions in the interval $[0, D]$, and ensure that the inter-antenna spacing is no less than $D_0$, respectively.

Note that problem~\eqref{op:pri_prob} is highly non-convex with respect to the design parameters $(\bw,\bt)$ since the objective and the constraint \eqref{eq:const2} are non-convex. Plus, the design of FA positions amounts to the optimization of the unit circle manifold. These properties make problem~\eqref{op:pri_prob} difficult to solve.

To achieve an efficient design algorithm for problem \eqref{op:pri_prob}, we take advantage of the block variable structure and extend the block successive upper bound minimization (BSUM) method~\cite{shi2011wmmse,shao2017simple,Yang2020BCD} to handle it. 
The unique challenge is to incorporate FA positions in the development. 
We outline the key steps of the tailored algorithm as follows.
\begin{equation}\label{op:WMMSE}
	\begin{split}
			\min_{ \bw, \bt, \bm \bu, \bm \rho }~
			{\cal F}(\bw, \bt, \bm \bu, \bm \rho)
			~~\mbox{s.t. }
			\setC_{\sf BS},~\setC_{\sf FR},~\setC_{\sf AC},~\setC_t,
		\end{split}
\end{equation}
where
$
{\cal F}(\bw, \bt, \bm \bu, \bm \rho) = \sum_{k=1}^{K} [\bw_k^{\rm H} \bA(\bt) \bw_k - 2\mathscr{R}\{\bb_k^{\rm H}(\bt)\bw_k\}  - {\rm log}(\rho_k) ]
$
with $\bu\in \Cbb^{K}$ and $\bm\rho\in \Rbb_{++}^K$ being introduced auxiliary variables; $\bA(\bt) = \sum_{k=1}^{K}\rho_k |u_k|^2 \bh_k(\bt)\bh_k^{\rm H}(\bt) + \sigma_k^2$; $\bb_k(\bt)=\rho_k u_k \bh_k(\bt)$; and $\mathscr{R}\{ x \}$ denoting the real part of $x$.
By leveraging the above transformation, we apply BSUM and obtain the following
\begin{subequations}\label{eq:BCD}
	\begin{align}
		\bu^{\ell+1} = &~\arg\min_{\bu \in \Cbb^{K}} {\cal F}(\bw^{\ell}, \bt^{\ell}, \bm \bu, \bm \rho^{\ell}); \label{subproblem_u} \\
		\bm \rho^{\ell+1} = &~\arg\min_{\bm \rho \in \Rbb_{++}^K}  {\cal F}(\bw^{\ell}, \bt^{\ell}, \bm \bu^{\ell+1}, \bm \rho); \label{subproblem_rho} \\
		\bw^{\ell+1} = &~\arg\min_{\bw \in \setC_{\sf BS} \cap \setC_{t}} {\cal F}(\bw, \bt^{\ell}, \bm \bu^{\ell+1}, \bm \rho^{\ell+1}); \label{subproblem_w} \\
		\bt^{\ell+1} = &~\arg\min_{\bt \in \setC_{t} \cap \setC_{\sf FR} \cap \setC_{\sf AC}} {\cal F}(\bw^{\ell+1}, \bt, \bm \bu^{\ell+1}, \bm \rho^{\ell+1}). \label{subproblem_x}
	\end{align}
\end{subequations}
Notably, the optimizations of $\bu$ in~\eqref{subproblem_u} and $\bm\rho$ in~\eqref{subproblem_rho} have closed-form solutions, which are given by
\begin{equation*}
	\begin{split}
		u_k^{\ell+1} = \frac{ \bh_k^{\rm H} \bw_k  }{\sum_{i=1}^{K} |\bh_k^{\rm H} \bw_i|^2 + \sigma_k^2 }, ~
		\rho_k^{\ell+1} = \frac{1}{1 - (u_k^{\ell+1})^* \bh_k^{\rm H} \bw_k }
	\end{split}
\end{equation*}
for any $k \in {\cal K}$; and $(\cdot)^*$ denotes the complex conjugate.

However, the optimizations of the beamformer $\bw$ and the APV $\bt$ do not have closed forms. 
We custom-build efficient algorithms for these two subproblems.

\vspace{-0.5cm}
\subsection{Beamformer Optimization}
The subproblems \eqref{subproblem_w} involve coupled constraints, and the projections onto the coupled constraints can be difficult to calculate. 
To tackle this challenge, we propose a PDA that can effectively decouple the constraints and yield efficient closed-form updates.
By PDA, we approximate \eqref{subproblem_w} by
\begin{equation} \label{op:pri_proj}
	\begin{split}
		\min_{ \bw }~
		 {\cal F}(\bw)  + \bar\rho \, {\rm dist}^2(\bw, \setC_{\sf BS}) + \bar\rho\, {\rm dist}^2(\bw, \setC_t),
	\end{split}
\end{equation}
where $\mbox{dist}(\bw, \setX)$ denotes the distance from point $\bw$ to set $\setX$;
$\bar\rho>0$ is a given penalty parameter. Obviously, the optimal solution of problem~\eqref{op:pri_proj} is equivalent to the solution of problem~\eqref{subproblem_w} when $\bar\rho$ is large.

Still, the distance function is variational and does not have an explicit expression.  
We address this issue by considering majorizating them by
\begin{equation}\label{eq:dis_maj_w}
	\mbox{dist}(\bw, \setC_{\sf BS}) \leq \|\bw - \tw_{\sf BS}\|_2,~
	\mbox{dist}(\bw, \setC_{t}) \leq \|\bw - \tw_{t}\|_2,
\end{equation}
where $\tw_{\sf BS} = \Pi_{{\cal C}_{\sf BS}}(\bw)$; $\tw_{t} = \Pi_{{\cal C}_{t}}(\bw)$; and the notation
\[
\Pi_{{\cal C}_{{\cal X}}}(\bw) = \arg\min_{\by \in {\cal X}} \| \by-\bw \|_2^2
\]
denotes projecting $\bw$ onto $\setX$. By meticulous calculation, we show that
\begin{equation*}
	\begin{split}
		[\tw_{\sf BS}]_k = &\begin{cases}
			\bw_k, & \mbox{if } \| \bw \|_2^2 \leq P_{max}, \\
			\sqrt{P_{max}}\frac{\bw_k}{\| \bw \|_2} , & \mbox{otherwise},
		\end{cases}\\
	\end{split}
\end{equation*}
\begin{equation*}
	\begin{split}
		[\tw_{t}]_k = &\begin{cases}
			\bw_k, ~~~ {\rm if }~ \sum_{k=1}^{K} \bw_k^{\rm H} \bda(\bt,\theta) \bda^{\rm H}(\bt,\theta) \bw_k \geq P_t, \\
			\left(\bI - \mu \bda(\bt,\theta) \bda^{\rm H}(\bt,\theta)\right)^{-1} \bw_k , ~~~ {\rm otherwise},
		\end{cases}
	\end{split}
\end{equation*}
where $\mu = \frac{1}{ \Vert \bda(\bt,\theta) \Vert_2^2} - \sqrt{ \frac{\sum_{k=1}^{K} \bw_k^{\rm H} \bda(\bt,\theta) \bda^{\rm H}(\bt,\theta) \bw_k}{|\bda^{\rm H}(\bt,\theta) \bda(\bt,\theta)|^2 P_t}  }.$
The derivation of the projection $\tw_{t}$ is shown in Appendix A.

By the distance majorization, at iteration $\bw$, one only needs to solve an unconstrained quadratic programming problem, given by
\begin{equation} \label{op:pda}
	\begin{split}
		\min_{ \bw }~
		{\cal F}(\bw)  + \bar\rho \left( \|\bw - \tw_{\sf BS}\|_2^2 + \|\bw - \tw_{t}\|_2^2 \right),
	\end{split}
\end{equation}
which has an optimal closed-form solution, i.e.,
\begin{equation}
	\begin{split}
		\bw_k = \left(\bA + 2 \bar\rho\bI \right)^{-1} \left[ \bar\rho \left([\tw_{\sf BS}]_k + [\tw_{t}]_k \right) + \bb_k \right],\, k\in {\cal K}.
	\end{split}
\end{equation}
The PDA was summarized in Algorithm 1. Note that we also apply the extrapolation in step 3, which was shown to be able to numerically accelerate the algorithmic convergence~\cite{Li2020PDA}.
\begin{algorithm}[t]
	\caption{An Efficient PDA for Problem \eqref{subproblem_w}}
	\begin{algorithmic}[1]
		\STATE {\bf Input:}  Initialize $\bar\rho > 0$, $\kappa>1$, $i = 1$, $\bw_k^{0} = \bw_k^{1}$, $k\in{\cal K}$.
		\STATE {\bf Repeat:} 		
		\STATE  \qquad $\bz_k^{i} = \bw_k^{i} + \frac{i-1}{i+2} (\bw_k^{i} - \bw_k^{i - 1})$;		
		\STATE  \qquad $\by_k^{i} = \Pi_{{\cal C}_{\sf BS}}(\bz_k^{i}) + \Pi_{{\cal C}_t}(\bz_k^{i})$;	
		\STATE  \qquad $\bw_k^{i+1} = \left(\bA + 2 \bar\rho\bI \right)^{-1} \left( \bar\rho\by_k^{i} + \bb_k \right)$;	
		\STATE  \qquad  $i = i + 1$;
		\STATE  \qquad  $\bar\rho = \kappa \bar\rho$ every $I$ iterations;
		\STATE {\bf Until} stopping criterion is satisfied.
	\end{algorithmic}
\end{algorithm}

\subsection{APV Optimization}
An EPG algorithm is proposed for handling problem~\eqref{subproblem_x}, which takes the following updates
\begin{equation} \label{gradient_method}
	\begin{split}
		& \bt^{i + 1} = \Pi_{\setC_t \cap \setC_{\sf FR} \cap \setC_{\sf AC}} \left( \bz^{i} - \eta \nabla_{\bt} {\cal F}(\bz^{i} | \bt^{i}) \right), \\
		& \bz^{i+1} = \bt^{i+1} + \zeta_{i+1}( \bt^{i+1} - \bt^{i} ),
	\end{split}
\end{equation}
where $\eta>0$ is the descent step length, which is obtained by the backtracking line search;
$\nabla_{\bt} {\cal F}(\bz^{i} | \bt^{i})$ denotes the gradient of ${\cal F}$ at $\bz^{i}$, and
the parameter $\zeta_{i+1} = \frac{\alpha_{i+1} - 1}{\alpha_{i+1}}$, $\alpha_{i+1} = \frac{1+\sqrt{1+4\alpha_{i}^2}}{2}$ with $\alpha_{1} = 0$.

Defining $\bw_k = \ba_k + j \bb_k$, $\bC_k = \ba_k\ba_k^{\rm T} + \bb_k\bb_k^{\rm T}$, $\bD_k = \ba_k\bb_k^{\rm T} - \bb_k\ba_k^{\rm T}$, $\bg_k = [g_{1,k},g_{2,k},\dots,g_{M,k}]^{\rm T}$, $\bq_k = [q_{1,k},q_{2,k},\dots,q_{M,k}]^{\rm T}$, $g_{m,k} = {\rm cos}(v_k t_m)$, and $q_{m,k}={\rm sin}(v_k t_m)$ for any $k\in {\cal K}$, we have
\begin{equation*}
	\begin{split}
		f_{k,i} &= |\bda^{\rm H}(\bt,\theta_k) \bw_i|^2 = \bg_k^{\rm T} \bC_i \bg_k + \bq_k^{\rm T} \bC_i \bq_k + 2 \bg_k^{\rm T} \bD_i \bq_k,\\
		h_{k,k} &= \mathscr{R}\{\bda^{\rm H}(\bt,\theta_k) \bw_k \} = \bg_k^{\rm T} \ba_k + \bq_k^{\rm T}\bb_k.
	\end{split}
\end{equation*}
Thus, the derivatives of $f_{k,i}$ and $h_{k,k}$ with respect to $\bt$ gives
\begin{equation*}
	\begin{split}
		&\frac{\partial f_{k,i}}{\partial \bt} = \frac{\partial f_{k,i}}{\partial \bg_k} \frac{\partial \bg_k}{\partial \bt} + \frac{\partial f_{k,i}}{\partial \bq_k} \frac{\partial \bq_k}{\partial \bt} \\
		&= 2v_k \left[\diag(\bg_k)( \bC_i\bq_k - \bD_i\bg_k ) - \diag(\bq_k)( \bC_i\bg_k + \bD_i\bq_k ) \right], \\
		&\frac{\partial h_{k,k}}{\partial \bt} = \frac{\partial h_{k,k}}{\partial \bg_k} \frac{\partial \bg_k}{\partial \bt} + \frac{\partial h_{k,k}}{\partial \bq_k} \frac{\partial \bq_k}{\partial \bt} \\
		&= v_k\left[ \diag(\bg_k) \bb_k - \diag(\bq_k) \ba_k \right].
	\end{split}
\end{equation*}
Next, $\nabla_{\bt} {\cal F}(\bt)$ can be derived and is given by
\begin{equation*}
	\begin{split}
		\nabla_{\bt} {\cal F}(\bt) = \sum_{k=1}^{K} \left\{ \rho_k |u_k|^2 \delta_k^2 \sum_{i=1}^{K} \frac{\partial f_{k,i}}{\partial \bt} - 
		 2\mathscr{R}\{\rho_k u_k^* \delta_k \}\frac{\partial h_{k,k}}{\partial \bt}\right\}.
	\end{split}
\end{equation*}

Moreover, $\Pi_{\setC_t \cap \setC_{\sf FR} \cap \setC_{\sf AC}}(\bm\kappa)$ can be obtained by solving the following problem
\begin{equation} \label{project_x}
	\begin{split}
		\min_{ \bt }~
		\Vert \bt - \bm\kappa \Vert_2^2
		~~~\mbox{s.t. }~ \bt \in
		\setC_t \cap \setC_{\sf FR}\cap\setC_{\sf AC},
	\end{split}
\end{equation}
where $\bm\kappa$ is a point to project.
However, the constraint $\setC_t$ is non-convex.
Therefore, we construct a concave quadratic surrogate function to minorize $\bda^{\rm H}(\bt,\theta) \bR_w \bda(\bt,\theta) $ as
\begin{equation} \label{global_lower_bound}
	\begin{split}
		\bda^{\rm H}(\bt,\theta) \bR_w \bda(\bt,\theta) \geq g(\bt|\tilde\bt) \triangleq \bt^{\rm T} \bD \bt - 2 \bd^{\rm T} \bt + c,
	\end{split}
\end{equation}
where $\bD = -v^2 \left(\diag(\br) - \bR\right)$,
\[
\bd[n] = v\sum_{m=1}^{M} |R_{mn}| {\rm sin}(f(\tilde t_n,\tilde t_m)) -v^2\sum_{m=1}^{M} |R_{mn}| (\tilde t_n - \tilde t_m),
\]
\begin{equation*}
	\begin{split}
		c = &\sum_{m=1}^{M} \sum_{n=1}^{M} |R_{mn}|[ {\rm cos}(f(\tilde t_n,\tilde t_m)) + v{\rm sin}(f(\tilde t_n,\tilde t_m)) (\tilde t_n - \tilde t_m) \\ & - \frac{1}{2}v^2 (\tilde t_n - \tilde t_m)^2 ],
	\end{split}
\end{equation*}
with $\br = \left[\sum_{m=1}^{M}|R_{m1}|,\sum_{m=1}^{M}|R_{m2}|,\dots,\sum_{m=1}^{M}|R_{mM}|  \right]$; $v = \frac{2\pi}{\lambda} {\rm cos}(\theta)$; $[\bR]_{mn} = |R_{mn}|$; $f(\tilde t_n,\tilde t_m) = v(\tilde t_n-\tilde t_m) + \angle R_{mn}$; $\tilde\bt = [\tilde t_1,\tilde t_2,\dots,\tilde t_M]^{\rm T}$; $R_{mn}$ is an element of the $m$th row and $n$th column of $\bR_w$ for $n,m = 1,2,\dots, M$; and $\tilde\bt$ denotes any determined value for $\bt$.

Therefore, the problem~\eqref{project_x} is reformulated as follows
\begin{equation} \label{convex_x}
	\begin{split}
		\min_{ \bt }~
		\Vert \bt - \bm\kappa \Vert_2^2
		~~\mbox{s.t. }~
		\bt \in
		\setC_q \cap \setC_{\sf FR}\cap\setC_{\sf AC},
	\end{split}
\end{equation}
where the feasible space $\setC_q$ of $\bt$ is restricted by $g(\bt|\bt^i)\geq P_t$.

Problem~\eqref{convex_x} is a convex quadratically constrained quadratic program (QCQP) problem, which can be solved efficiently with off-the-shelf toolboxes, such as CVX. 

\vspace{-0.25cm}
\subsection{Computational Complexity Analysis}
In this subsection, we analyze the complexity of the proposed BSUM algorithm in \eqref{eq:BCD}.
The computational complexity of the BSUM algorithm is ${\cal O}(M^{4.5} + M^2(MK+M+K))$.
This is because the overall complexity of the closed-form $\bu$, as well as $\bm\rho$, is ${\cal O}(M^2K)$, the complexity for beamformer optimization \eqref{subproblem_w} is ${\cal O}(M^3 + M^2K)$, and the complexity for APV optimization \eqref{subproblem_x} is ${\cal O}(M^{4.5} + M^3K)$.

\vspace{-0.25cm}
\section{Simulation Results}
\begin{figure}[t]
	\centering
	\subfloat[Performance Comparison]{\includegraphics[width=0.46\columnwidth]{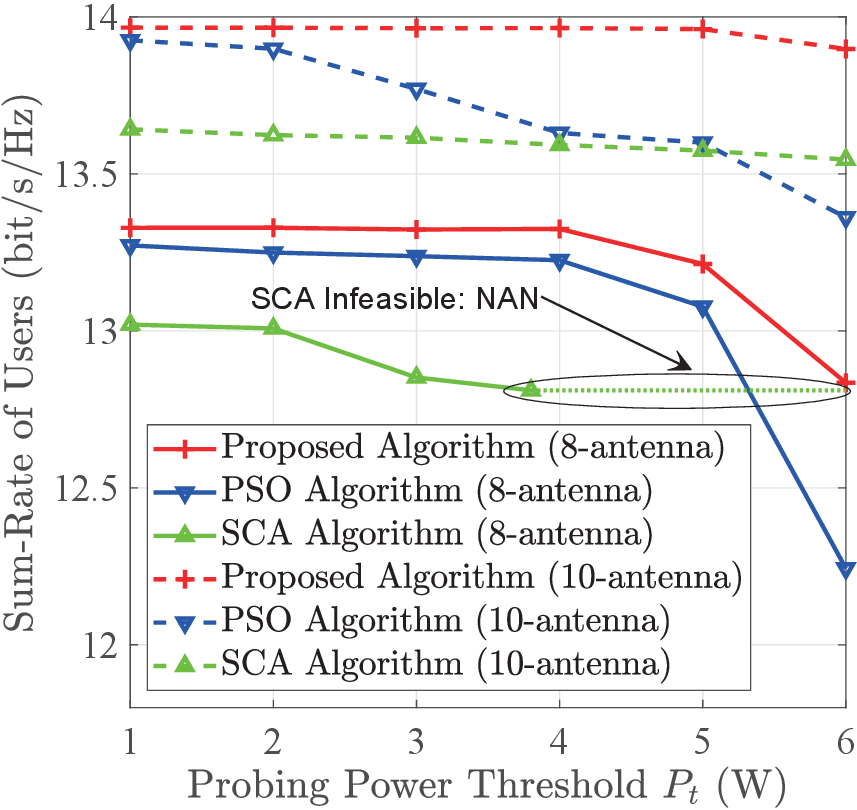}%
		\label{algorithm_performance}}
	\hfil
	\subfloat[Average Runtimes ]{\includegraphics[width=0.455\columnwidth]{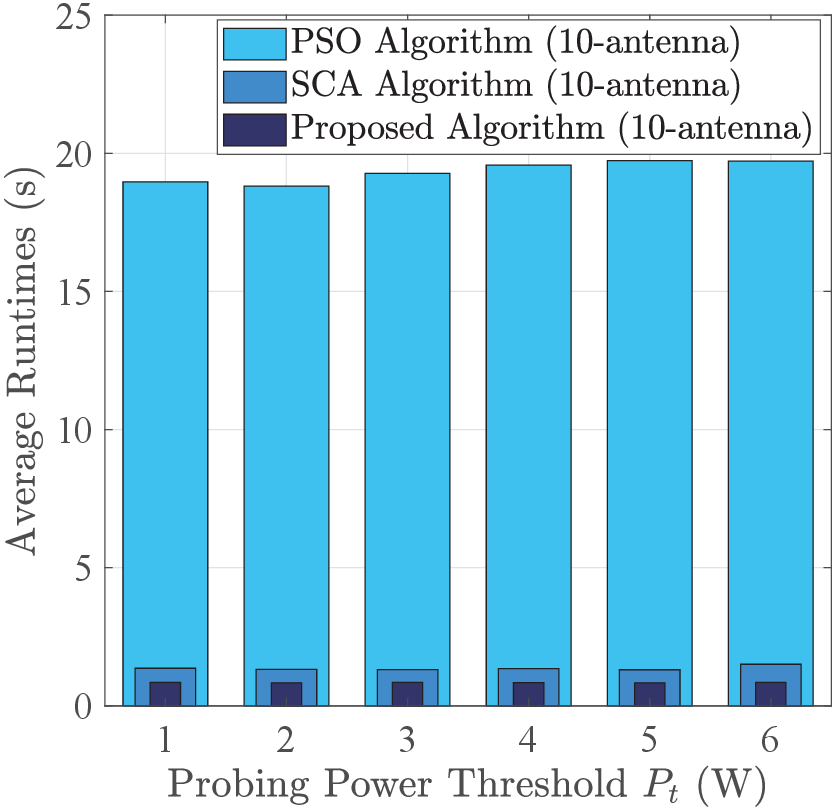}%
		\label{algorithm_average_runtimes}}
	\hfil
	\caption{Comparison between the proposed algorithm and existing algorithms with different $P_t$. $K=2$, $P_{max}=30\,$dBm.}
	\label{fig:algorithm_comparison}
	\vspace{-0.5cm}
\end{figure}
\begin{figure}[t]
	\centering
	\subfloat[Performance Comparison]{\includegraphics[width=0.46\columnwidth]{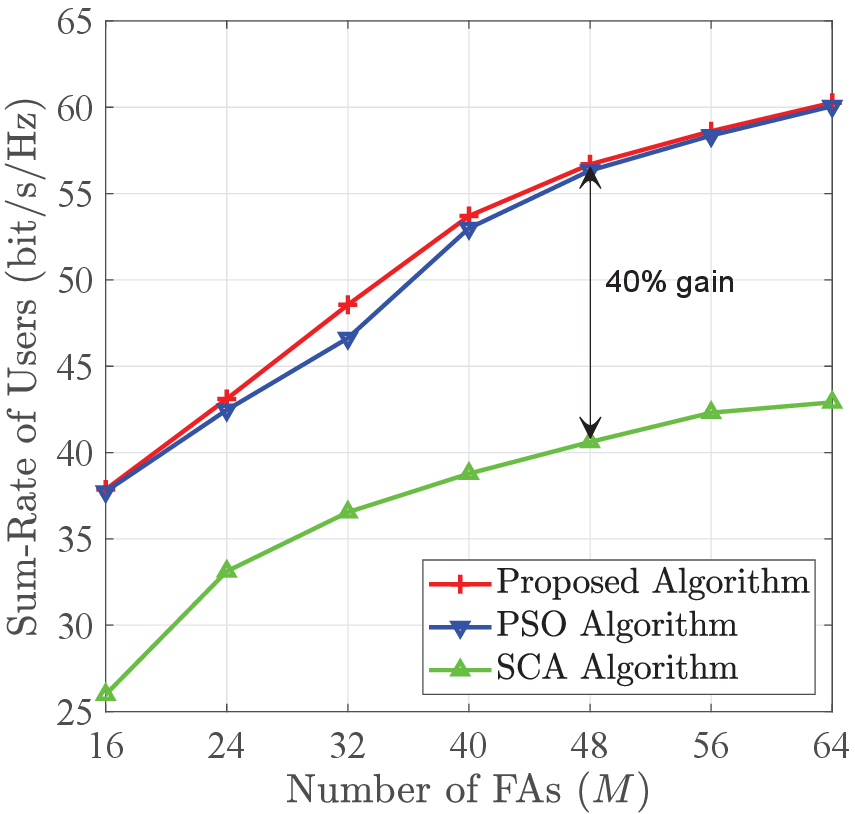}%
		\label{algorithm_performance_M}}
	\hfil
	\subfloat[Average Runtimes ]{\includegraphics[width=0.468\columnwidth]{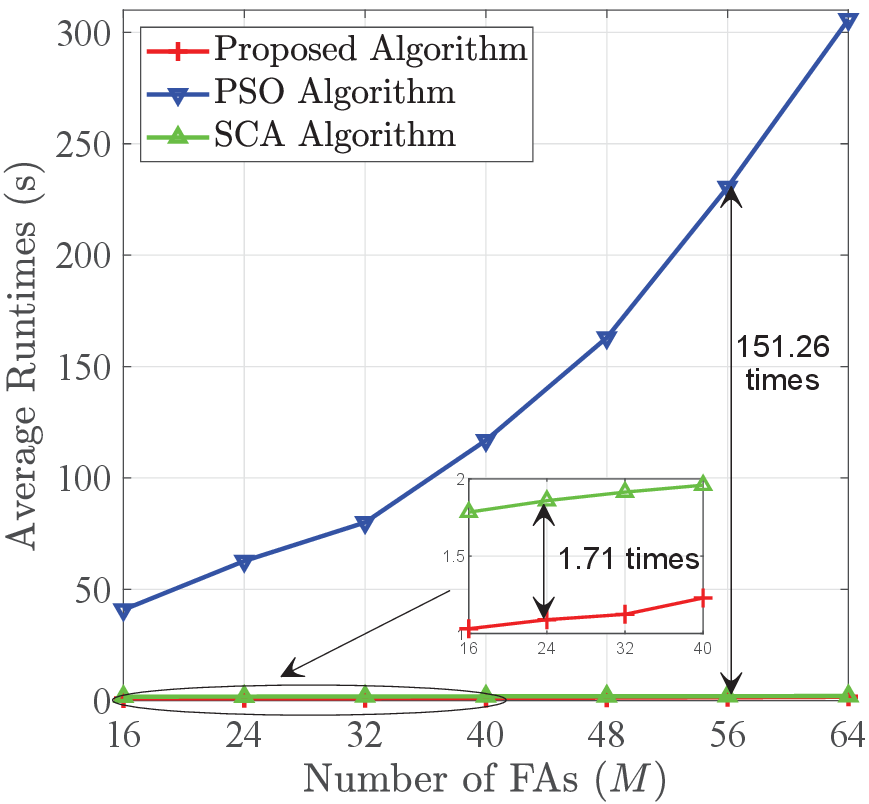}%
		\label{algorithm_runtime_M}}
	\hfil
	\caption{Comparison between the proposed algorithm and existing algorithms with different $M$. $K=8$, $D=M\lambda$, $P_{max}=30\,$dBm, $P_t=6\,$W.}
	\label{fig:algorithm}
	\vspace{-0.55cm}
\end{figure}
\begin{figure}[t]
	\centering
	\vspace{-0.05cm}
	\subfloat[Underloaded $K=2$]{\includegraphics[width=0.46\columnwidth]{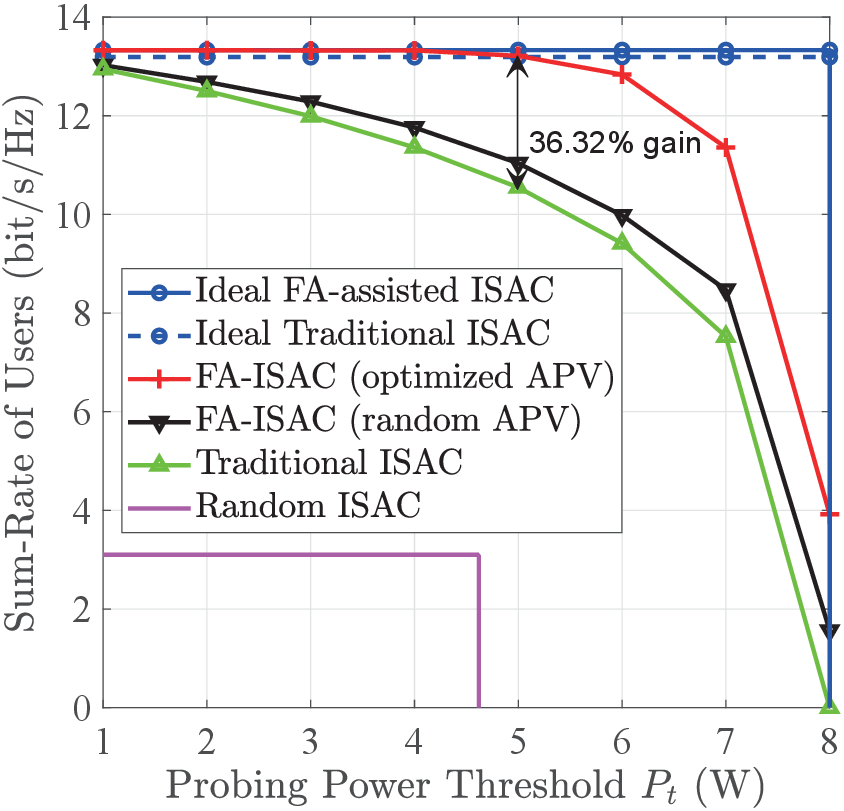}%
		\label{trade_off_8_antenna_2_user}}
	\hfil
	\subfloat[Overloaded $K=8$]{\includegraphics[width=0.46\columnwidth]{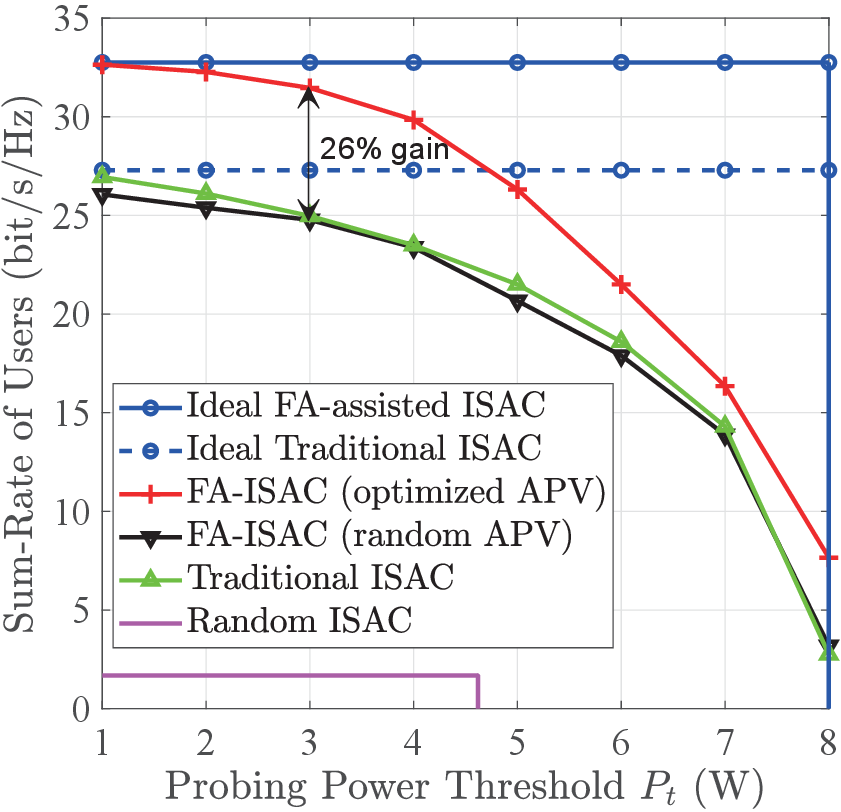}%
		\label{trade_off_8_antenna_8_user}}
	\hfil
	\caption{Trade-off performance between multiuser sum rate and probing power. $M=8,~P_{max}=30\,$dBm.}
	\vspace{-0.55cm}
	\label{fig:performance_trade_off}
\end{figure}
\begin{figure}[t]
	\centering
	\subfloat[Multiuser Sum-Rate]{\includegraphics[width=0.46\columnwidth]{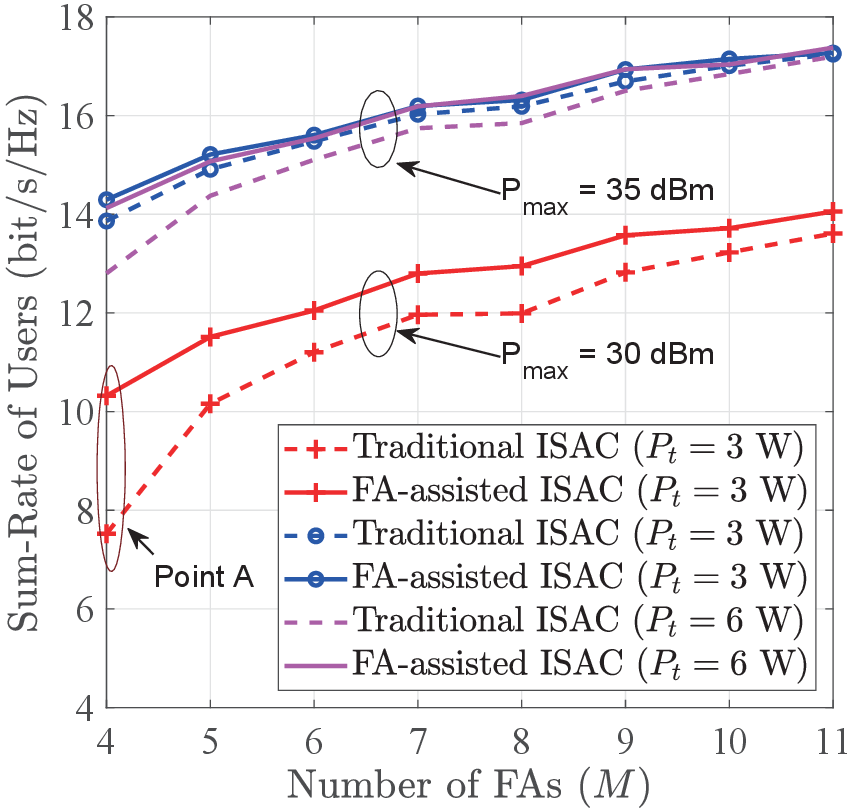}%
		\label{sum_rate_different_M}}
	\hfil
	\subfloat[Beampattern at Point A of (a)]{\includegraphics[width=0.455\columnwidth]{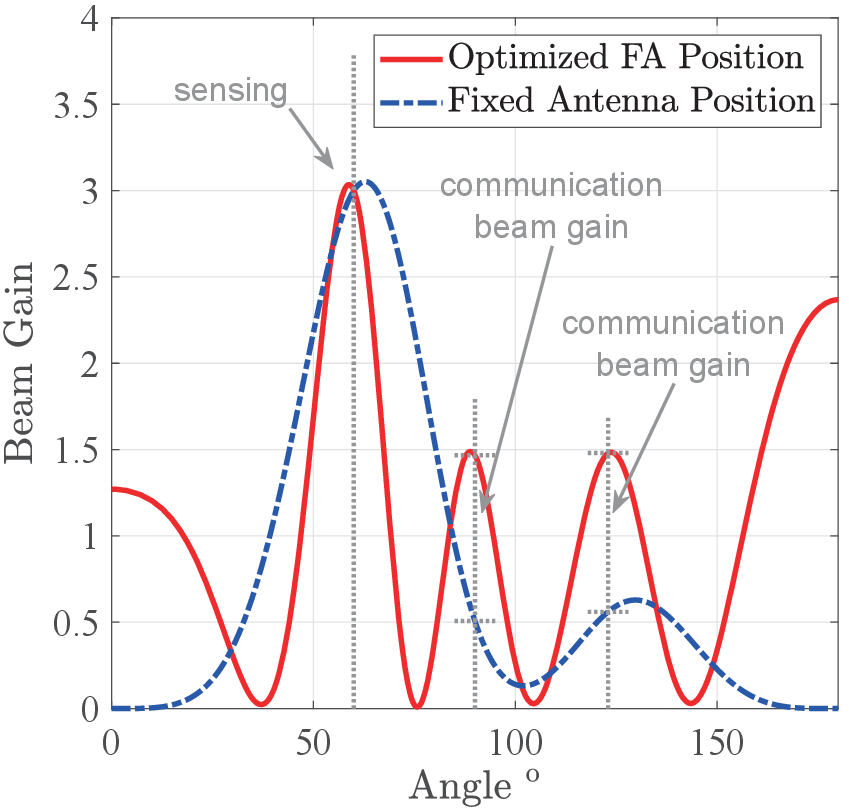}%
		\label{beam_gain}}
	\hfil
	\caption{Multiuser sum rate under different number of BS antenna. $K=2$.}
	\vspace{-0.5cm}
	\label{fig:sum_rate_beampattern}
\end{figure}
In this section, the simulations demonstrate the proposed BSUM algorithm performance in FA-assisted ISAC systems.
We set the parameters of FAs as $D_0 = \lambda/2$, $D = 10\lambda$, and $\lambda = 0.01\,$m~\cite{Ma2024MA,Zhu2024MA_Modeling}.
In addition, we set $\theta = 60^{\circ}$, $\sigma_k^2 = -80\,$dBm, and $\delta_k = g_0 d_k^{-\alpha}$ for any $k\in {\cal K}$, where $g_0 = -40\,$dB and $\alpha = 2.8$ denote the fading at $1\,$m reference distance and the path-loss exponent, respectively.
Users are distributed on a circle centered at BS with radius 100m, i.e., $d_k=100\,$m for $k \in {\cal K} $.
In this section, we consider two scenarios, i.e, the 2-user (underloaded) scenario and the 8-user (overloaded) scenario.
For the 2-user scenario, we set the AOD of the FA array corresponding to two users to be $\theta_1 = 90^{\circ}$ and $\theta_2 = 120^{\circ}$, respectively.
For the 8-antenna scenario, we set the AOD of the FA array corresponding to eight users to be $\theta_1 = 10^{\circ}$, $\theta_2 = 30^{\circ}$, $\theta_3 = 80^{\circ}$, $\theta_4 = 90^{\circ}$, $\theta_5 = 120^{\circ}$, $\theta_6 = 130^{\circ}$, $\theta_7 = 150^{\circ}$, and $\theta_8 = 170^{\circ}$, respectively.

As shown in Fig.~\ref{fig:algorithm_comparison}, we compare the commonly used algorithms in FA system design, i.e., the SCA algorithm \cite{Wu2024MA_ISAC,Ma2024MA} and the PSO algorithm \cite{Kuang2024MA_ISAC,Zuo2024FA_PSO}.
The proposed algorithm has a higher performance compared to the SCA algorithm and PSO algorithm and can improve the speed of FA position configuration by 60\% compared to the SCA algorithm, and by more than 20 times compared to the PSO algorithm.
In particular, when $M= 8$, the SCA algorithm causes problem~\eqref{eq:BCD} to be infeasible as the probing power increases, due to the fact that SCA reduces the feasible region.
Further, from Fig.~\ref{fig:algorithm}, the proposed algorithm can obtain more than $40\%$ performance gain and save 70\% runtimes than SCA algorithm.
In addition, the proposed algorithm is able to operate more than 150 times faster while obtaining slightly better performance than the PSO algorithm.

Fig.~\ref{fig:performance_trade_off} shows the performance trade-off between multiuser sum rate and probing power under the underloaded scenario and the overloaded scenario.
Among, ideal ISAC performance is obtained by maximizing the multiuser sum rate in the ISAC system when $P_t=0$ is set.
From Fig.~\ref{fig:performance_trade_off}, under the low sensing requirements in underloaded scenarios, the FPA can provide enough DoF to serve users, and thus the FA array has a low enhancement of ISAC performance. 
However, in overload scenarios, the FPA no longer has sufficient capability to serve users, while FA can greatly improve ISAC performance by providing higher space DoF. 
Under the high sensing requirements, FAs achieve a marked improvement in ISAC performance compared to the FPA in underloaded scenarios, which is because the FPA array cannot well balance high probing power and multiuser communications.
In overloaded scenarios, the FA still has sufficient capacity to balance sensing and communication, providing considerable performance gain than the FPA, as indicated in Fig.~\ref{fig:performance_trade_off}$\,$(b).

Fig.~\ref{fig:sum_rate_beampattern}$\,$(a) shows that the performance difference between the FA array and the FPA array is larger when $M$ is small, e.g., $M\in [4,7]$.
This is due to the fact that the smaller $M$ means a larger positional space achieving a higher DoF gain.
Furthermore, when $P_t = 3\,$W, the performance difference between FA and FPA arrays decreases with increasing $P_{max}$ due to the fact that the larger power budget provides higher power DoF.
When $P_{max}=35\,$dBm, the performance difference between FA and FPA arrays increases with the increase of $P_t$. 
This is because the higher probing power makes the FPA array not sufficiently capable enough to provide better service to users.
Finally, we give the beampattern for point A in Fig.~\ref{fig:sum_rate_beampattern}$\,$(a), as shown in Fig.~\ref{fig:sum_rate_beampattern}$\,$(b).
It can be seen that the FPA array is unable to provide the communication beam in the desired direction while guaranteeing the sensing requirement. 
In contrast, the communication beam in the desired direction can be achieved by optimizing the FA position, which provides a higher communication gain compared to the FPA.

\vspace{-0.045cm}
\section{Conclusion}
This letter proposed an efficient algorithm for an FA-assisted ISAC system, aiming at maximizing the multiuser sum rates with the required sensing requirement.
Relying on the MM algorithm, the PDA was proposed to iteratively update the closed-form beamformer, and the EPG was presented to efficiently configure the FA positions.
Simulation results demonstrated that the proposed algorithm can realize FA position configuration more than 60\% faster than the SCA and PSO algorithms.

\vspace{-0.078cm}
\begin{appendices}
	\section{}
	$\tw_{t}$ can be derived by solving the following problem
	\begin{equation} \label{op:projection}
		\min_{ \bw }~
		\Vert \bw - \bm\xi \Vert_2^2
		\quad\mbox{s.t. }~
		\sum_{k=1}^{K} \bw_k^{\rm H} \bda\bda^{\rm H} \bw_k \geq P_t,
	\end{equation}
	where $\bm\xi = [\bm\xi_1^{\rm T},\bm\xi_2^{\rm T},\dots,\bm\xi_K^{\rm T}]^{\rm T}$ is the point to project, and $\bda$ denotes $\bda(\bt,\theta)$ for short.

	The Karush–Kuhn–Tucker (KKT) conditions associated with the above problem are given by
	\begin{equation*}
		\begin{split}
			& \frac{\partial {\cal L}}{\partial \bw_k} = 2(\bw_k - \bm\xi_k) - 2\mu \bda \bda^{\rm H} \bw_k = 0,~k \in {\cal K}, \\
			& \sum_{k=1}^{K} \bw_k^{\rm H} \bda \bda^{\rm H} \bw_k \geq P_t,~\mu \geq 0,~
			\mu \big( \sum_{k=1}^{K}\bw_k^{\rm H} \bda \bda^{\rm H} \bw_k - P_t \big) = 0,
		\end{split}
	\end{equation*}
where   $\mu \geq 0$ is a dual variable.
	From the first line of KKT condition and Woodbury matrix identity, we get
	\begin{equation} \label{solution}
		\begin{split}
			\bw_k^{\star} = (\bI - \mu \bda \bda^{\rm H})^{-1} \bm\xi_k = \left( \bI + \frac{\mu \bda\bda^{\rm H}}{1 - \mu \bda^{\rm H} \bda} \right) \bm\xi_k,
		\end{split}
	\end{equation}
By the complementary slackness, we know that if $\sum_{k=1}^{K} \bw_k^{\rm H} \bda \bda^{\rm H} \bw_k > P_t$, then $\mu = 0$ must hold.
	Otherwise, $\mu$ is chosen such that $\sum_{k=1}^{K} \bw_k^{\rm H} \bda \bda^{\rm H} \bw_k = P_t$, given by
	\begin{equation}
		\begin{split}
			\mu = \frac{1}{\Vert \bda \Vert^2} - \sqrt{ \frac{\sum_{k=1}^{K} \bm\xi_k^{\rm H} \bda \bda^{\rm H} \bm\xi_k}{|\bda^{\rm H}\bda|^2 P_t}  }.
		\end{split}
	\end{equation}
	
\end{appendices}

\bibliographystyle{IEEEtran}
\bibliography{refs}

\begin{thebibliography}{10}
\providecommand{\url}[1]{#1}
\csname url@samestyle\endcsname
\providecommand{\newblock}{\relax}
\providecommand{\bibinfo}[2]{#2}
\providecommand{\BIBentrySTDinterwordspacing}{\spaceskip=0pt\relax}
\providecommand{\BIBentryALTinterwordstretchfactor}{4}
\providecommand{\BIBentryALTinterwordspacing}{\spaceskip=\fontdimen2\font plus
\BIBentryALTinterwordstretchfactor\fontdimen3\font minus
  \fontdimen4\font\relax}
\providecommand{\BIBforeignlanguage}[2]{{%
\expandafter\ifx\csname l@#1\endcsname\relax
\typeout{** WARNING: IEEEtran.bst: No hyphenation pattern has been}%
\typeout{** loaded for the language `#1'. Using the pattern for}%
\typeout{** the default language instead.}%
\else
\language=\csname l@#1\endcsname
\fi
#2}}
\providecommand{\BIBdecl}{\relax}
\BIBdecl

\bibitem{Cui2021ISAC}
Y.~Cui, F.~Liu, X.~Jing, and J.~Mu, ``Integrating sensing and communications
  for ubiquitous {IoT}: Applications, trends, and challenges,'' \emph{IEEE
  Netw.}, vol.~35, no.~5, p. 158–167, Sep. 2021.

\bibitem{Liu2018ISAC}
F.~Liu, C.~Masouros, A.~Li, H.~Sun, and L.~Hanzo, ``{MU-MIMO} communications
  with {MIMO} radar: From co-existence to joint transmission,'' \emph{IEEE
  Trans. Wireless Commun.}, vol.~17, no.~4, pp. 2755--2770, Apr. 2018.

\bibitem{Dong2023R_ISAC}
Y.~Dong, F.~Liu, and Y.~Xiong, ``Joint receiver design for integrated sensing
  and communications,'' \emph{IEEE Commun. Lett.}, vol.~27, no.~7, pp.
  1854--1858, Jul. 2023.

\bibitem{Liu2018DFRC}
F.~Liu, L.~Zhou, C.~Masouros, A.~Li, W.~Luo, and A.~Petropulu, ``Toward
  dual-functional radar-communication systems: Optimal waveform design,''
  \emph{IEEE Trans. Signal Process.}, vol.~66, no.~16, pp. 4264--4279, Aug.
  2018.

\bibitem{Liu2022ISAC}
F.~Liu, {Y. -F. Liu}, A.~Li, C.~Masouros, and Y.~C. Eldar, ``{Cramér-Rao}
  bound optimization for joint radar-communication beamforming,'' \emph{IEEE
  Trans. Signal Process.}, vol.~70, p. 240–253, Jan. 2022.

\bibitem{Zhang2023ISAC_EI}
T.~Zhang, G.~Li, S.~Wang, G.~Zhu, G.~Chen, and R.~Wang, ``{ISAC}-accelerated
  edge intelligence: Framework, optimization, and analysis,'' \emph{IEEE Trans.
  Green Commun. Netw.}, vol.~7, no.~1, pp. 455--468, Mar. 2023.

\bibitem{Wong2023FA}
{K. -K. Wong}, W.~K. New, X.~Hao, {K. -F. Tong}, and {C. -B. Chae}, ``Fluid
  antenna system—{Part I}: Preliminaries,'' \emph{IEEE Commun. Lett.},
  vol.~27, no.~8, pp. 1919--1923, Aug. 2023.

\bibitem{Xiao2024CEMA}
L.~Z. Y. L. B. N.~X. Z.~Xiao, S.~Cao and R.~Zhang, ``Channel estimation for
  movable antenna communication systems: A framework based on compressed
  sensing,'' \emph{IEEE Trans. Wireless Commun.}, vol.~23, no.~9, pp.
  11\,814--11\,830, Sept. 2024.

\bibitem{New2024FAChannelEstimation_2}
W.~K.~N. {\it et al.}, ``Channel estimation and reconstruction in fluid antenna
  system: Oversampling is essential,'' \emph{arXiv preprint arXiv:2405.15607},
  2024.

\bibitem{Wang2024FA_ISAC}
C.~Wang, G.~Li, H.~Zhang, {K. -K. Wong}, Z.~Li, D.~W.~K. Ng, and {C. -B. Chae},
  ``Fluid antenna system liberating multiuser {MIMO} for {ISAC} via deep
  reinforcement learning,'' \emph{IEEE Trans. Wireless Commun.}, vol.~23,
  no.~9, pp. 10\,879--10\,894, Sept. 2024.

\bibitem{Zou2024FA_ISAC}
J.~Zou, H.~Xu, C.~Wang, L.~Xu, S.~Sun, K.~Meng, C.~Masouros, and {K. -K. Wong},
  ``Shifting the {ISAC} trade-off with fluid antenna systems,'' \emph{arXiv
  preprint arXiv:2405.05715}, 2024.

\bibitem{Kuang2024MA_ISAC}
Z.~Kuang, W.~Liu, C.~Wang, Z.~Jin, J.~Ren, X.~Zhang, and Y.~Shen,
  ``Movable-antenna array empowered {ISAC} systems for low-altitude economy,''
  \emph{arXiv preprint arXiv:2406.07374}, 2024.

\bibitem{Wu2024MA_ISAC}
H.~Wu, H.~Ren, and C.~Pan, ``Movable antenna-enabled {RIS}-aided integrated
  sensing and communication,'' \emph{arXiv preprint arXiv:2407.03228}, 2024.

\bibitem{Wong2023FA_PartII}
{K. -K. Wong}, {K. -F. Tong}, and {C. -B. Chae}, ``Fluid antenna system—part
  {II}: Research opportunities,'' \emph{IEEE Commun. Lett.}, vol.~27, no.~8,
  pp. 1924--1928, Aug. 2023.

\bibitem{Stoica2007Radar}
P.~Stoica, J.~Li, and Y.~Xie, ``On probing signal design for {MIMO} radar,''
  \emph{IEEE Trans. Signal Process.}, vol.~55, no.~8, p. 4151–4161, Aug.
  2007.

\bibitem{shi2011wmmse}
Q.~Shi, M.~Razaviyayn, {Z. -Q. Luo}, and C.~He, ``An iteratively weighted
  {MMSE} approach to distributed sum-utility maximization for a {MIMO}
  interfering broadcast channel,'' \emph{IEEE Trans. Signal Process.}, vol.~59,
  no.~9, pp. 4331--4340, Aug. 2011.

\bibitem{shao2017simple}
M.~Shao and {W. -K. Ma}, ``A simple way to approximate average robust multiuser
  {MISO} transmit optimization under covariance-based {CSIT},'' in \emph{IEEE
  Int. Conf. Acoustics, Speech Signal Process. (ICASSP)}.\hskip 1em plus 0.5em
  minus 0.4em\relax New Orleans, LA, USA, 2017, pp. 3504--3508.

\bibitem{Yang2020BCD}
Z.~Q.~L. Y.~Yang, M.~Pesavento and B.~Ottersten, ``Inexact block coordinate
  descent algorithms for nonsmooth nonconvex optimization,'' \emph{IEEE Trans.
  Signal Process.}, vol.~68, pp. 947--961, 2020.

\bibitem{Li2020PDA}
Q.~Li, Y.~Liu, M.~Shao, and {W. -K. Ma}, ``Proximal distance algorithm for
  nonconvex {QCQP} with beamforming applications,'' in \emph{Proc. IEEE Int.
  Conf. Acoust., Speech, Signal Process. (ICASSP)}.\hskip 1em plus 0.5em minus
  0.4em\relax Barcelona, Spain, 2020, pp. 5155--5159.

\bibitem{Ma2024MA}
W.~Ma, L.~Zhu, and R.~Zhang, ``Multi-beam forming with movable-antenna array,''
  \emph{IEEE Commun. Lett.}, vol.~28, no.~3, pp. 697--701, Mar. 2024.

\bibitem{Zhu2024MA_Modeling}
L.~Zhu, W.~Ma, and R.~Zhang, ``Modeling and performance analysis for movable
  antenna enabled wireless communications,'' \emph{IEEE Trans. Wireless
  Commun.}, vol.~23, no.~6, pp. 6234--6250, Jun. 2024.

\bibitem{Zuo2024FA_PSO}
Y.~Zuo, J.~Guo, B.~Sheng, C.~Dai, F.~Xiao, and S.~Jin, ``Fluid antenna for
  mobile edge computing,'' \emph{IEEE Commun. Lett.}, vol.~28, no.~7, pp.
  1728--1732, Jul. 2024.

\end{thebibliography}

\end{document}